\documentclass[a4paper,10pt,twocolumn,article,oneside]{memoir} 
\usepackage[english]{babel}
\usepackage[utf8]{inputenc}
\usepackage[T1]{fontenc}	

\usepackage{authblk}
\usepackage{enumitem}
\usepackage{graphicx, graphbox} 
\usepackage{float}              

\usepackage{xcolor}
\usepackage{hyperref} 
\hypersetup{
    colorlinks,
    linkcolor={red!50!black},
    citecolor={blue!50!black},
    urlcolor={blue!80!black}
}
\usepackage[font=small,labelfont=bf,textfont=it]{caption} 

\usepackage{braket}
\usepackage{amsmath, amssymb, amsthm, mathtools}
\usepackage{cleveref}

\usepackage{tikz}
    \usetikzlibrary{calc}

\usepackage{amsthm}

\usepackage{csquotes}  
\usepackage[style=numeric-comp, autocite=inline, giveninits=true, sorting=none, url=false, isbn=false]{biblatex}
    \DeclareNameAlias{labelname}{given-family}  
\newcommand{\tensor}{\otimes}
\newcommand{\bigtensor}{\bigotimes}
\newcommand{\for}{\text{ for }}
\newcommand{\with}{\text{ with }}

\newcommand{\neighb}{\mathcal{N}}

\newcommand{\measure}{\mathcal{M}}
\newcommand{\cliff}{C}

\newcommand{\byprod}{\Sigma}

\DeclareMathOperator{\CZ}{CZ}
\DeclareMathOperator{\CX}{CX}

\DeclareMathOperator{\LC}{LC}
\usepackage{xparse}
\makeatletter
\newcounter{my_warnings}
\DeclareDocumentCommand{\warn}{g}{%
    \def\@message{\IfNoValueTF{#1}{WARNING!}{#1}}
    \colorbox{red}{\color{white}\MakeUppercase{\@message}}
    \stepcounter{my_warnings}
}
\makeatother

\title{\large\bfseries Applications and resource reductions in measurement-based variational quantum eigensolvers}

\author[1,2,*]{Frederik Kofoed Marqversen}
\author[1,2,$\dagger$]{Nikolaj Thomas Zinner}
\affil[1]{Department of Astronomy and Physics, Aarhus University, Denmark}
\affil[2]{Kvantify Aps, DK-2300 Copenhagen S, Denmark}
\affil[*]{marqversen@phys.au.dk}
\affil[$\dagger$]{zinner@phys.au.dk}

\date{\small (Dated: \today)}

\begin{document}
\twocolumn[
    \begin{@twocolumnfalse}
        \maketitle
        \renewcommand{\abstractname}{}
        \begin{abstract}
            We discuss the procedure for obtaining measurement-based implementations of quantum algorithms given by quantum circuit diagrams and how to reduce the required resources needed for a given measurement-based computation. This forms the foundation for quantum computing on photonic systems in the near term. To demonstrate that these ideas are well grounded we present three different problems which are solved by employing a measurement-based implementation of the variational quantum eigensolver algorithm (MBVQE). We show that by utilising native measurement-based gates rather than standard gates, such as the standard CNOT, MBQCs may be obtained that are both shallow and have simple connectivity while simultaneously exhibiting a large expressibility. We conclude that MBVQE has promising prospects for resource states that are not far from what is already available today.
        \end{abstract}
        \vspace{1em}
    \end{@twocolumnfalse}
]

\section*{Introduction}

Photonic systems have over recent years been proving to be a useful platform for demonstrating quantum advantage \cite{zhong_quantum_2020, madsen_quantum_2022}. However, current hardware structures are still limited in that they are restricted to Gaussian boson sampling \cite{kruse_detailed_2019} which is not a universal computational procedure. To open up the full potential of quantum computation, fully universal structures must be developed. One possible angle of attack is to consider measurement-based quantum computational (MBQC) methods like that first introduced by \citeauthor{raussendorf_measurement-based_2003} \cite{raussendorf_one-way_2001, raussendorf_measurement-based_2003}. 

Hybrid quantum-classical algorithms such as the variational quantum eigensolver (VQE), are of major interest as they show very promising results with a relatively small demand of resources and are more likely to be realisable for NISQ era devices \cite{peruzzo_variational_2014, kandala_hardware-efficient_2017, farhi_quantum_2014, tilly_variational_2022}. The idea to implement such hybrid quantum algorithms as MBQCs was first investigated by \citeauthor{ferguson_measurement-based_2021} \cite{ferguson_measurement-based_2021}. They suggested that the VQE algorithms developed in the circuit picture may directly be compiled into MBQCs. In MBQC, one may enact the Gottesman-Knill theorem through graph manipulations and thus reduce the size of such an implementation. However, we show that the reduced graphs from such a procedure might have undesirable geometry for practical use. Since the number of qubits scales with the depth of the algorithm such an approach will almost certainly grow out of what is feasibly realisable in experiment. This is particularly important for continuous variable MBQC, since finite squeezing limits the number of possible measurements \cite{alexander_noise_2014, larsen_fault-tolerant_2021}. For these reasons, it is imperative to tailor the VQE algorithm to MBQC.

In this article, \cref{sec:theory} is an introduction to the theory behind MBQC. The discussion gives a general overview of the main ideas and issues concerning measurement-based algorithms. All of the ideas are further distilled into \cref{sec:implementation} in which we discuss how these are put together to obtain a full quantum circuit to MBQC compiler. To investigate specific MBQC gates we use a tensor network structure to simulate the given MBQCs. The question of how this is done is the contents of \cref{sec:simulation}. In \cref{sec:MBVQE} we present a specifically measurement-based implementation of a layered VQE, as well as the numerical results from simulating the computation as applied to three distinct and relevant problems: Finding molecular ground states, determining the ground state of two-dimensional Heisenberg models, and solving the vehicle routing problem.
\section{Theoretical background}\label{sec:theory}

Measurement-based computations, in short MBQCs, are performed on some initially highly entangled states from which qubits are measured one by one. Measuring a single qubit unentangles that qubit from any other qubit it may have been entangled to. The qubit is effectively removed from the system, thus reducing the initial quantum state. In this way, the initial state is a resource that is being used up. Furthermore, quantum teleportation shows us that the action of measurement may have a non-trivial effect on the left-over qubits \cite{gottesman_demonstrating_1999}. Hence, measurements constitutes an operation, and a number of consecutive measurements a computation, an MBQC. In this section we discuss how the action of a given measurement-based gate is determined, how one deals with the non-deterministic nature of quantum measurements, how to reduce the size of the required resource state for a given computation, and finally we discuss the equivalence that exists between resource states that have the same computational potential.

The resource states used for MBQC are so-called \emph{graph states} \cite{hein_multiparty_2004, raussendorf_one-way_2001, raussendorf_measurement-based_2003}. Graph states are a particular type of quantum state that are represented by a mathematical graph. Vertices represent individual qubits and edges represent two-qubit entanglement between qubits. Specifically, for a given graph $G=(V, E)$ the corresponding graph state is
\begin{equation}
    \ket{G} = \prod_{(a, b) \in E} \CZ_{(a, b)} \ket{+}^{\tensor V},
    \label{eq:graph_state}
\end{equation}
where $\CZ$ is the controlled phase gate and $\ket{+} = (\ket{0} + \ket{1})/\sqrt{2}$. Graph states are also \emph{stabiliser states}, described by the stabiliser generated by the operators
\begin{equation}
    K_a = X_a \prod_{b \in \neighb_a} Z_b
        \qquad\forall a \in V,
\end{equation}
where $\neighb_a$ is the set of neighbours to $a$ within the graph. MBQC is thus quite naturally described by the stabiliser formalism \cite{nielsen_quantum_2012, gottesman_heisenberg_1998, gottesman_stabilizer_1997}.

\Citeauthor{raussendorf_measurement-based_2003} \cite{raussendorf_measurement-based_2003} show how, given a particular graph $G$ and a set of predetermined measurement axes thereon, the implemented unitary operation $U$ can be deduced up to local Pauli rotations $U_\byprod$ called the byproduct. In an application of this method, one can show that the non-trivial gate $R_{Z^{\tensor n}}(\theta) = \exp[-i \frac{\theta}{2} Z^{\tensor n}]$ is very naturally implementable as an MBQC \cite{browne_one-way_2006} by a single one-qubit measurement.

Due to the non-deterministic nature of quantum measurements, the action of an MBQC inevitably depends on the outcomes the applied measurements. However, the dependence is completely encompassed by the byproduct $U_\byprod$. A crucial part of MBQC as a computational model is to handle this byproduct. This is thoroughly discussed in ref. \cite{raussendorf_measurement-based_2003}. The basic idea is to propagate the byproduct through the gate in one of two ways
\begin{align}
    & U U_\byprod = (U U_\byprod U^\dagger) U = U_\byprod' U,
    & \for U \in \cliff_n
        \\
    & U U_\byprod = U_\byprod (U_\byprod^\dagger U U_\byprod) = U_\byprod U',
    & \for U \notin \cliff_n
\end{align}
where $\cliff_n$ is the Clifford group on $n$ qubits \cite{nielsen_quantum_2012}.

The gate $U'$ above will in general depend on the measurement outcomes. However, due to the local nature of the byproduct, this effect can be cancelled by allowing for some of the single qubit measurements in our computation to be what is called \emph{adaptive}. Adaptive measurements are simply measurements that depend on previous measurement results. Having propagated the byproduct through to the left it may be handled by yet another set of adaptive measurements. If $\measure$ are the final out-measurements one wishes to apply to the output of the MBQC, then $\measure' = U_\byprod \measure U_\byprod^\dagger$ are the measurements one must perform to cancel the effect of the byproduct.


The difficulty in performing quantum computation on a circuit lies in the hardness of successfully applying multi-qubit gates. In MBQC there are no multi-qubit operations. Here, the difficulty lies entirely in the preparation of the graph state. Preparing high-quality entangled states, however, is no easy task either. Some groups have successfully obtained some smaller graph states \cite{larsen_deterministic_2021, larsen_fault-tolerant_2021, vigliar_error-protected_2021, thomas_efficient_2022}, although more work needs to be done in order to obtain high fidelity graphs of appreciable size. For that reason, it is important to investigate possible routines for reducing the graph state needed to implement a specific computation. It turns out that significant reduction is possible due to the \emph{Gottesman-Knill theorem} \cite{gottesman_heisenberg_1998}. The theorem states that computations consisting of Clifford operations can be efficiently simulated classically, and since computations in MBQC are performed by measurements, such a simulation will take care of a large portion of the measure-qubits hence reducing the required graph.

\begin{figure}
    \centering
    \includegraphics[width=0.9\linewidth]{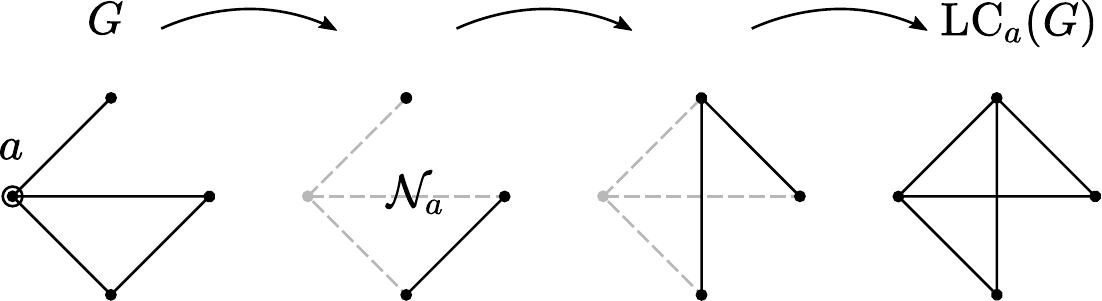}
    \caption{Illustrative example of local complementation of a simple four-vertex graph. The circled vertex $a$ is the vertex at which the local complementation is applied.}
    \label{fig:local_complementation}
\end{figure}

One way of realising the Gottesman-Knill theorem w.r.t. MBQC is by simulation of local (single-qubit) Pauli measurements through a graph manipulation called \emph{local complementation} which is illustrated in \cref{fig:local_complementation}. The local complement of a graph $G = (V, E)$ at a vertex $a \in V$ is obtained by complementing only the subgraph induced by the neighbourhood $\neighb_a$ of the vertex $a$. In ref. \cite{hein_multiparty_2004} it is shown that the graph state obtained from the local complement of $G$ is related to the graph state of $G$ itself by
\begin{gather}
    \ket{\LC_a(G)} = U^{\LC}_a \ket{G}
        ,\\
    \with U^{\LC}_a = \sqrt{+iX_a} \prod_{b \in \neighb_a} \sqrt{-iZ_b}.
\end{gather}
This result is called the LC-rule. 

The rule can be used to transform measurements in the Pauli $X$ and $Y$ bases into measurements in the Pauli $Z$ basis on an LC-equivalent graph. The term \emph{LC-equivalence} refer to the fact that the state that results from a measurement is not a proper graph state, but is still equivalent to one under the equivalence of multiplication of a local Clifford operator $\in (\cliff_1)^{\tensor n}$. Notice that graphs that are related through the LC-rule are in fact LC-equivalent. On top of that, one may show \cite{hein_multiparty_2004} that Pauli $Z$ measurements may be simulated by simply deleting the measured vertex from the graph. These rules constitute an efficient simulation of local Pauli measurements, thus reducing the size of the required graph state. Since all Clifford gates are implementable by Pauli measurements \cite{raussendorf_measurement-based_2003}, these simulations realise the Gottesman-Knill theorem on MBQCs.

\begin{figure}
    \centering
    \includegraphics[width=0.8\linewidth]{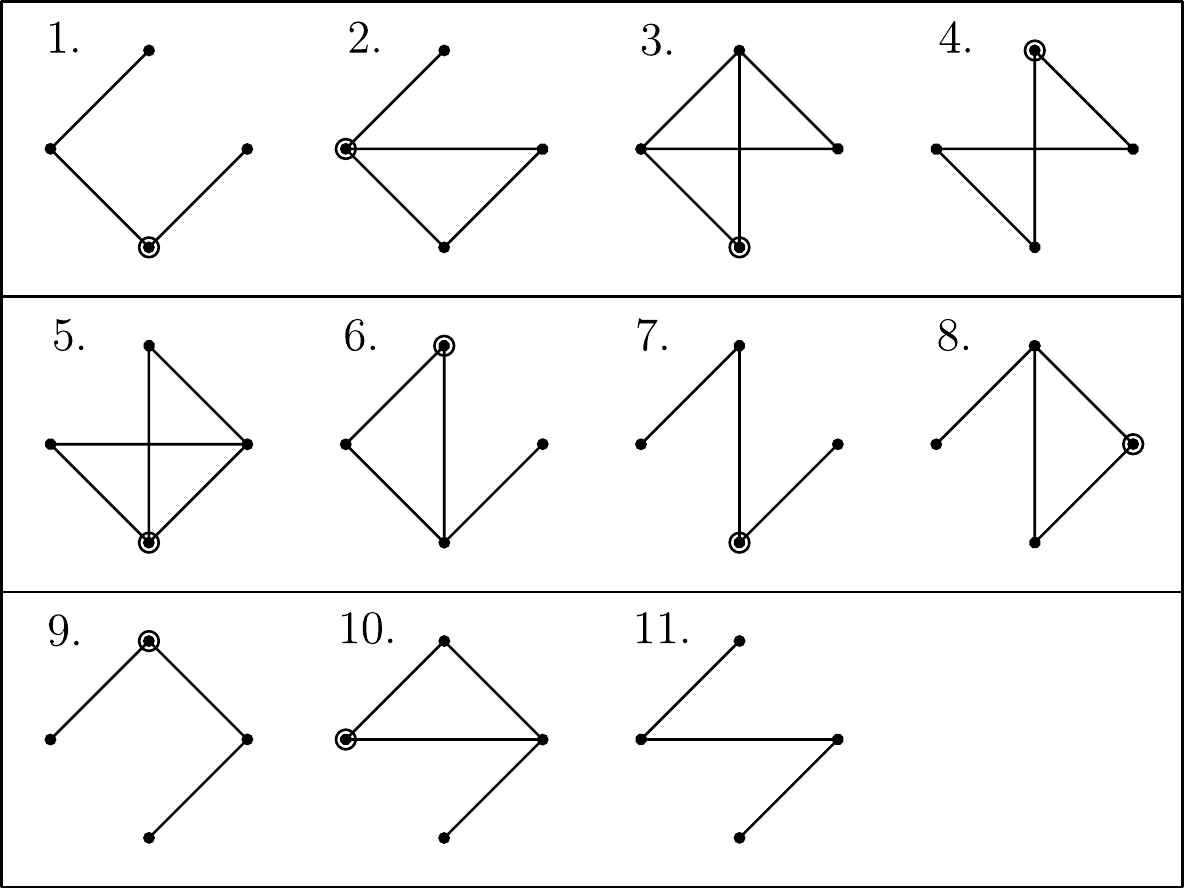}
    \caption{The full equivalence class of graph 1. The encircled vertices are those at which to apply local complementation to obtain the next. The figure is a reconstruction based on ref. \cite{hein_multiparty_2004}.}
    \label{fig:lc_equivalence_class}
\end{figure}

The advantage of the LC-equivalence relation is that any two graphs that are LC-equivalent also can be considered equivalent for MBQC. Two such graph states contain the same entanglement properties and thus may implement the same MBQCs simply by transforming the applied measurement pattern. \Cref{fig:lc_equivalence_class} depicts a cycle of 11 LC-equivalent graphs obtainable by sequentially applying local complementation \cite{hein_multiparty_2004}. If an experiment can realise a single one of those, then an MBQC implemented on any of the others can also be implemented on that one. This relaxes the restrictions put on hardware and opens up a range of possible computations given a specific graph state.
\section{Implementation overview}\label{sec:implementation}
An MBQC can be represented by an object which contains a graph. Each vertex contain the measurement basis it is to be measured in. The MBQC object will also contain two ordered lists of vertices that are labelled input and output qubits respectively. Finally, the object store information about the byproduct of the specific computation. The byproduct itself can be represented as two subsets of the vertices for each output qubit by the following mappings
\begin{gather}
    U_\byprod = \bigtensor_{i\in V_I} \left(Z^{\sum S_{z, i}} X^{\sum S_{x, i}} \right)
        ~\mapsto~
    \left[U_0, U_1, ...\right],
        \\
    U_i = Z^{\sum S_{z, i}} X^{\sum S_{x, i}} ~\mapsto~ \left[S_{z, i}, S_{x, i}\right],
\end{gather}
where $S_{x, i}$ is a subset of the measurement results which is mapped to the corresponding set of vertices. MBQC objects further will have a method that may propagate a byproduct through itself. The method differs depending on whether the gate is Clifford or not. All in all the method computes
\begin{gather}
    U U_\byprod ~\mapsto~ U_\byprod' U'.
\end{gather}
If the gate is not Clifford, this introduces adaptive measurements. The information on which qubits the measurement depends is stored together with the measurement bases at each vertex in the graph.

A factory for obtaining MBQC objects for several relevant gates must be built. MBQC implementations of a universal gate set and a few additional gates can be found in ref. \cite{raussendorf_measurement-based_2003}.

A circuit consists of a series of applied gates. To translate a circuit the MBQC class will thus need a method for concatenating two MBQC instances. This method must compute the series composition of the two graphs and then propagate the byproduct of the first through the second MBQC as follows:
\begin{equation}
    U_{\byprod_2}' U_2' U_{\byprod_1}' U_1' 
    = U_{\byprod_2}' (U_{\byprod_1}'' U_2'') U_1'
    = U_{\byprod_{tot}} U_2'' U_1'.
\end{equation}
With this propagation method, the MBQC objects must be able to handle \emph{any} byproduct and not just their own. An alternative method is to handle the byproduct of gate 1 by adapting the measurement bases of its output qubits, which are the same as the inputs for gate 2. With this alternative, there is no need for a generalised propagation scheme. This comes with the cost of adding additional adaptive measurements that might not be possible to simulate in the reduction process.

To include the reduction scheme with the compiler, the LC-rule must be implemented as a method. The LC-rule comprises two steps. First: Take the local complement of the graph stored in the MBQC
\begin{equation}
    G ~\mapsto~ \LC_a(G).
\end{equation}
second: Transform the measurement bases of the MBQC according to the applied local Clifford operator
\begin{equation}
    \measure ~\mapsto~ \measure' = U^{\LC}_a \measure (U^{\LC}_a)^\dagger.
    \label{eq:transformed_measurements}
\end{equation}
This transform applies to any output $X, Y, Z$ measurements as well.

Simulating the measuring of a qubit along a Pauli axis then simply amounts to applying the LC-rule as dictated by the results stated earlier. After simulating a measurement the measured vertex is left isolated and can be deleted. Since this constitutes a proper simulation, the measurement outcomes can be chosen arbitrarily. Fully Clifford reducing a given MBQC is performed by simulating all Pauli measurements within that MBQC.

One should address the order in which the Pauli measurements are simulated. The order turns out to have appreciable implications on the resulting graph. The graphs produced for different orderings will be LC-equivalent and thus will have the same computational capabilities. However, in most non-trivial cases, a random ordering will produce graphs with a large number of edges with have no apparent structure. The problem of finding the simplest graph within an LC-equivalence class has previously been addressed by \citeauthor{houshmand_minimal_2018} \cite{houshmand_minimal_2018} by introducing the notion of generalised flow (gflow) on measurement patterns, which has been implemented in a software solution \cite{sunami_graphix_2022} similar to that written and used by the authors of the present paper. For both methods, however, the resultant resource state is not guarantied to fit within a given physical architecture, and finding one that does is non-trivial. For that reason it is still important to investigate gates that are naturally simple in MBQC and has a known and fixed structure.
\section{Simulation of MBQC}\label{sec:simulation}
The size of the resource needed for a certain measurement-based computation is much larger than the input and output, and thus, classically storing the full state vector of the resource is exceedingly costly. Even with the Gottesman-Knill theorem, any interesting quantum computation will include a number of non-Clifford operations. This implies that even when simulating out the Clifford parts, the computation still require more qubits than that needed to simply store the output. For that reason, it is not preferable to simulate MBQCs as measurements on a state vector. The problem is remedied by the use of tensor networks as a simulation tool

\begin{figure}
    \centering
    \includegraphics[width=0.8\linewidth]{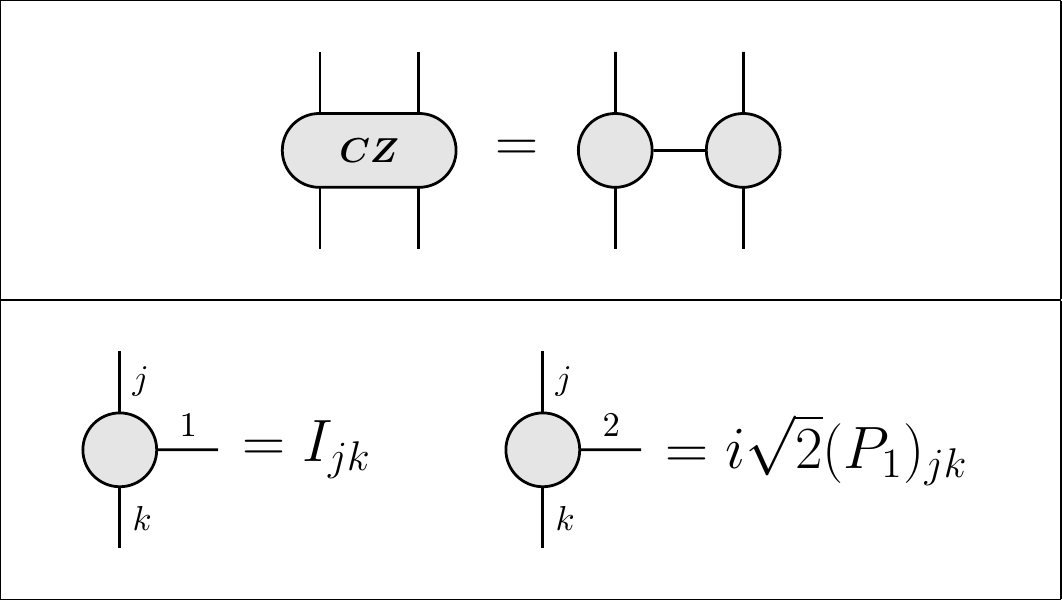}
    \caption{Symmetric tensor decomposition of the $\CZ$ operator. The two tensors in the decomposition are equal and defined as depicted in the bottom half of the figure. $I$ is the identity and $P_1 = \ket{1}\bra{1}$ the projection onto state 1}
    \label{fig:CZ_decomposition}
\end{figure}

It is well known that graph states can be represented as a \emph{projected entangled pair state} (PEPS) and that PEPS' are very naturally represented as \emph{tensor networks} \cite{orus_practical_2014}. The simplest way of obtaining a tensor network representation of a graph state is as follows: Construct the vector nodes corresponding to the initial state of all vertices. Then construct the edges one by one by applying the $\CZ$ operator between each neighbouring pair of vertices. To obtain the PEPS representation of the graph state one can use the tensor decomposition of the $\CZ$ operator shown in \cref{fig:CZ_decomposition}. The decomposition follows from the observation $\CZ = I \tensor I - 2 P_1 \tensor P_1$ where $I$ is the identity and $P_1 = \ket{1}\bra{1}$ the projection onto state 1.

The scaling of a tensor network (TN) is (number of parameters)
\begin{equation}
    m(\mathrm{TN}) = \smashoperator{\sum_{T \in \mathrm{TN}}} m(T) = \smashoperator[l]{\sum_{T \in \mathrm{TN}}} \smashoperator[r]{\prod_{e \in T}} \dim(e),
\end{equation}
where $e \in T$ represents all edges $e$ of the tensor-node $T$. The tensor network representation of a graph state is said to be efficient in that the memory required scales only polynomially in the number of qubits. Measurement of singular qubits is performed by projection onto the outcome eigenvector. In a tensor network, one-qubit state vectors are tensors with a single two-dimensional edge. Thus not only are graph states efficiently represented by tensor networks but so is an MBQC including the measurements.

When contracting an edge of a vector node, the tensor node it was connected to is relieved of two dimensions and its size reduced by a factor of two. Also, the vector node itself is deleted relieving the network of a further two parameters. The advantage of using tensor networks is that the order in which edges are contracted is arbitrary. Thus, measurements may be performed (by contracting the vector-node edges) without having to compute the graph state state-vector. An MBQC can be simulated in this way by always contracting those edges that result in a minimally costly network, significantly reducing the memory cost of the simulation. Unfortunately, the optimal way of contracting a tensor network is hard to determine, and thus clever methods of contracting a network must be employed.

The result of contracting a tensor network is a single node that represents the output state of the MBQC. This node still scales exponentially with the number of output qubits. However, the use of a tensor network reduces the simulation of an MBQC to be exponential in the output qubits, rather than in the total number of qubits in the MBQC. Computing the result of a quantum algorithm by simulating the MBQC with a tensor network is comparable to, sometimes faster than, computing the same algorithm by matrix multiplication onto the input state.

Simulating MBQCs using tensor networks is very natural and supplies the immediate optimisation discussed above. However, using tensor networks has further advantages, one of which is in computing expectation values of Hamiltonians, especially when approximations are allowed \cite{orus_practical_2014}. In the present work no such approximations are made, and expectation values are evaluated explicitly by tensor contraction (matrix multiplication).

Our simulations have been performed using the Python Library \textit{TensorNetwork} \cite{roberts_tensornetwork_2019}, and proceed as follows: First we build the PEPS representation of the graph for the particular MBQC we wish to compute. This is done as per the prescription in \cref{{eq:graph_state}} using the $CZ$ tensor decomposition from \cref{fig:CZ_decomposition}. Measurements are included in the tensor network each as a partial inner-product with the measurement-outcome state. The result is still a PEPS representation. The whole tensor network is contracted using a simple greedy algorithm, and the resulting final node will have dangling edges, and represents the output state of the MBQC. This state is then used in computing expectation values of the problem Hamiltonian where the MBQC byproduct is handled as per the prescription in \cref{eq:transformed_measurements}.
\section{Measurement-based VQE (MBVQE)}\label{sec:MBVQE}
In ref. \cite{ferguson_measurement-based_2021}, \citeauthor{ferguson_measurement-based_2021} discuss measurement-based variational methods. In this section, we argue that in order to obtain interesting results from small and simple resource states one must tailor algorithms from blocks that are natively measurement-based. In particular we show how this may be done with the VQE algorithm with only relatively light requirements on required resources.

\begin{figure}
    \centering
    \includegraphics[width=0.8\linewidth]{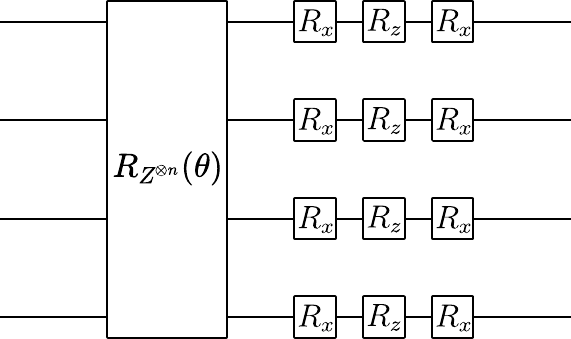}
    
    \vspace{0.5cm}
    
    \includegraphics[width=0.7\linewidth]{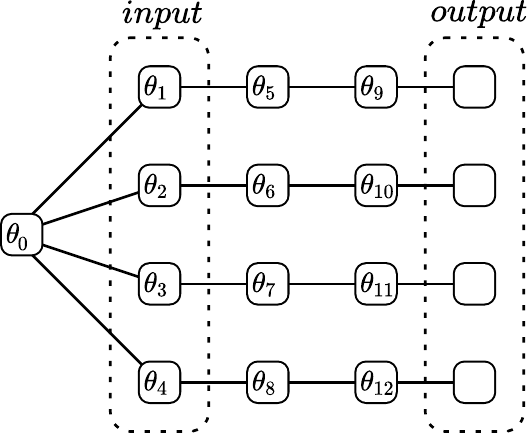}
    \hspace{0.8cm}
    
    \caption{(Top) Circuit for one layer of a hardware efficient VQE ansatz. It consists of a parametrised entangling step followed by a general Euler rotation of each qubit. The entangling step consist of a single application of the $n$-fold $Z$-rotation gate $R_{Z^{\tensor n}}(\theta) = \exp[-i \frac{\theta}{2} Z^{\tensor n}]$. (Bottom) The reduced MBQC that implements the circuit. Each box represents a physical qubit and the angles represent the rotated bases they are to be measured in.}
    \label{fig:vqe_circuit}
\end{figure}

For a layered VQE, one starts by arbitrarily rotating the initial qubits, after which, a sequence of alike layers is applied \cite{liu_layer_2022}. However, the dependencies for both angles and byproduct are independent of the numerical value of the parameters. We thus compute the MBQC that implements a layer of the ansatz and may then compute the total VQE by concatenation of layers depending on how many are needed for a specific task. The circuit that we have chosen for the layers of the VQE is shown in the topmost half of \cref{fig:vqe_circuit}. A layer thus consists of an entangling step followed by a general Euler rotation of each qubit. As the entangling gate we chose the diagonal $R_{Z^{\tensor n}}$ since it naturally implements as an MBQC, reducing the resources. Also, there is evidence that parametrising the entangling step of a VQE reduces required depths further reducing the resources \cite{rasmussen_parameterized_2022}. The bottom half of \cref{fig:vqe_circuit} shows the reduced MBQC that implements the single layer circuit. All the measurements are in general adaptive and the angles cannot be determined before knowing the outcome of preceding measurements. They are, however, equal to the rotational angles in the gate-based circuit up to a change of sign. The number of qubits in the MBQC equals the number of parameters in each VQE-layer plus $3n$ needed for initial rotations and an additional $n$ for storing the output state. That is 
\begin{gather}
    \text{\#PARAM} = d(3n + 1) + 3n
        \\
    \text{\#QUBITS} = \text{\#PARAM} + n
    \label{eq:qubit_count}
\end{gather}
where $d$ is the number of layers, also called the depth, of the ansatz, and $n$ the problem size.

The authors of ref. \cite{ferguson_measurement-based_2021} suggest that measurement-based implementations of well-studied algorithms can be obtained by compiling the circuit-based quantum algorithm into an MBQC and then be realised as such. This is true only if resources are not limited. In the NISQ era, however, where reduction is a necessity, more care must be taken in order to construct simple and realisable MBQCs.

\begin{figure}
    \centering
    \includegraphics[width=0.55\linewidth]{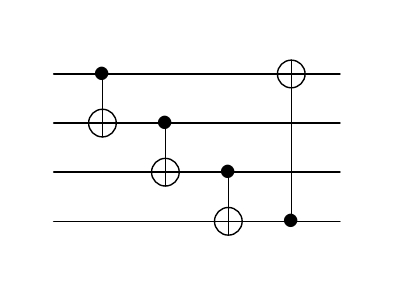}
    \includegraphics[width=0.4\linewidth]{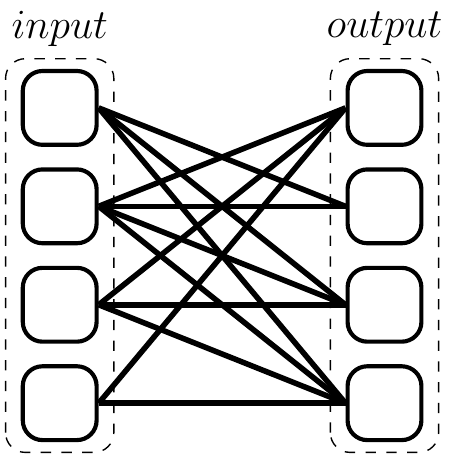}
    
    \caption{(Left) Quantum circuit with cylindrical nearest neighbour interaction through CX gates. (Right) A reduced MBQC implementation of the circuit from the left half of the figure, obtained by concatenation of measurement-based CNOT gates followed by Pauli measurement simulation as discussed in \cref{sec:theory} in a random order. The edges of the graph may be further simplified through LC-equivalence, although finding such a transformation is far from trivial.}
    \label{fig:4cx}
\end{figure}

Composite entangling gates can be reduced by using the previously discussed procedure just as any other MBQC. In the case of the standard controlled not CX, which is Clifford, the whole thing may be reduced out of the computation. The reduction procedure involves local complementation which introduces edges in a highly non-trivial fashion, and the result of the reduction procedure often results in graphs with a lot of edges between far apart qubits, and may not even exhibit any apparent symmetry reminiscent of the original MBQC, see \cref{fig:4cx}. These kinds of graphs are unattractive since they are likely to be too hard to realistically realise in experiment. Although possibly LC-equivalent to a simple graph, determining the simpler one is intractable for large graphs. Using $R_{Z^{\tensor 4}}$ has none of these problems. On top of that, since the input and output qubits are the same for these gates, this is true even for entangling gates that are concatenations of several $R_{Z^{\tensor 4}}$-type gates. This is the main reason for studying this particular type of gate.

For all instances in the following sections, we have employed the COBYLA algorithm for minimising the energy function within the VQE algorithm. This choice has been taken as it is a constrained parameter and gradient free method.

\subsection*{Molecular ground states}
\begin{figure}
    \centering
    \includegraphics[width=\linewidth]{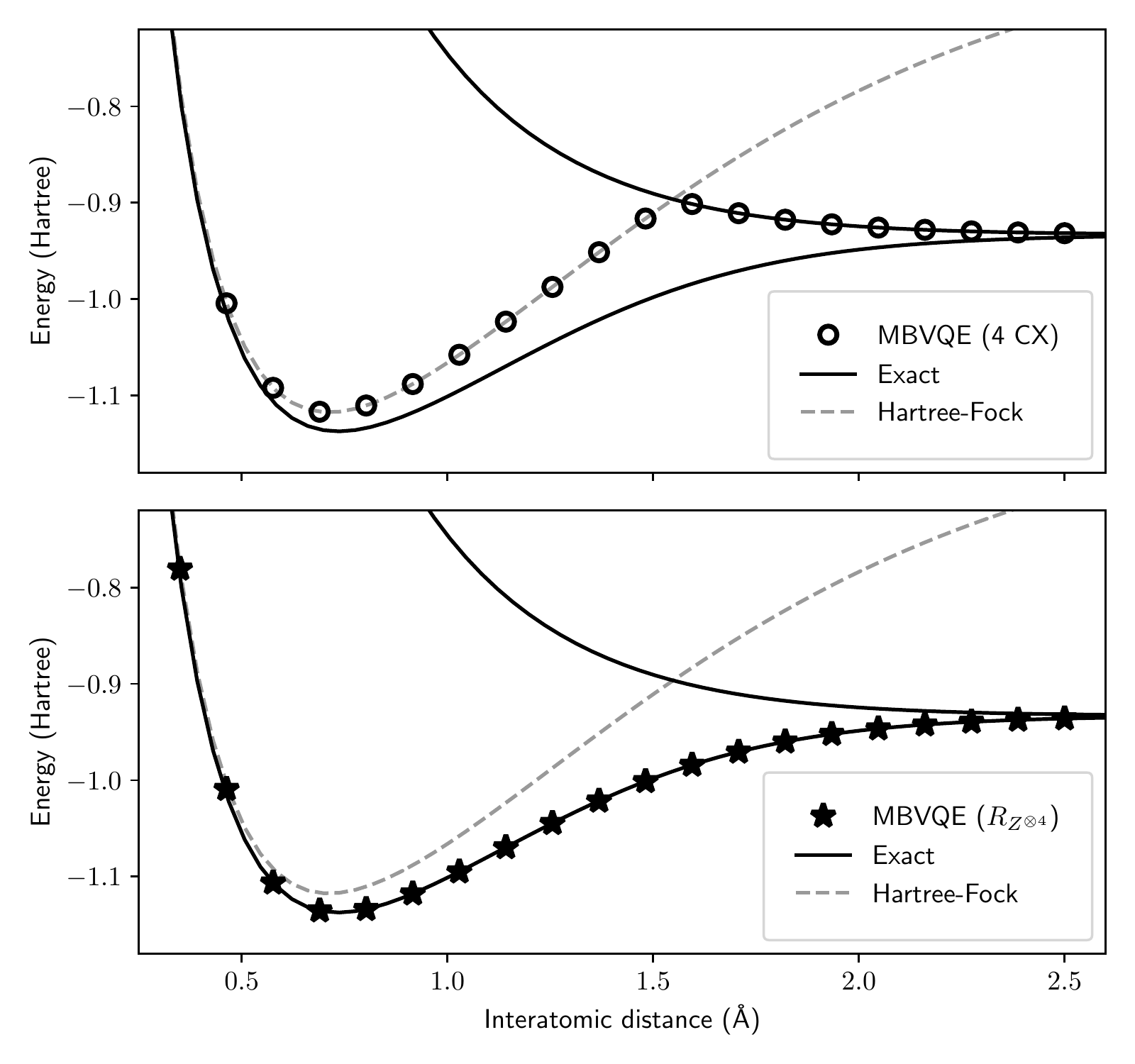}
    \caption{MBVQE results to the ground state energy of the $\text{H}_2$ molecule Hamiltonian in the STO-3G basis. The top figure contains the results using a static entangling gate consisting of 4 $\CX$ gates in a cylindrical pattern and the bottom figure the results using the parametrised entangling gate $R_{Z^{\tensor n}}(\theta)$ which is native to MBQC. The VQE algorithm was run for 20 equidistant interatomic distances in both cases.}
    \label{fig:h2}
\end{figure}

One of the chemical problems that has been used extensively as a standard benchmark for VQE is determining the ground state of the $\text{H}_2$ molecule \cite{kandala_hardware-efficient_2017}. The qubit Hamiltonian for the molecule is obtained by applying the Bravyi-Kitaev encoding to the STO-3G minimal basis representation of the $\text{H}_2$ Hamiltonian. \Cref{fig:h2} contain the results of simulating the MBVQE algorithm on the $\text{H}_2$ Hamiltonian at 10 different inter-atomic distances. For the $\text{H}_2$ molecule it has already been illustrated that when using simple and static entangling steps in the VQE, a single layer results in so-called \emph{kinks} in the results \cite{ryabinkin_constrained_2018}. This is what is shown in the top half of \cref{fig:h2} where the entangling gate is a standard CNOT between nearest neighbours as shown in \cref{fig:4cx}. Since this type of entangling takes no variational parameter the number of parameters are the 3 initial rotations and 3 final rotations of each of the 4 qubits. That is $4 \times 6 = 24$ parameters. Correspondingly the reduced MBVQE require $24+4 = 28$ qubits, taking into account the 4 qubits holding the output state.

It is known that the kink in the results can be fixed by increasing the depth to 2 layers \cite{ryabinkin_constrained_2018}. However, as is evident from the bottom half of \cref{fig:h2}, a single layer is enough when the entangling step is $R_{Z^{\tensor 4}}$. This type of entangling takes a single parameter and thus only require a single qubit in the MBVQE implementation per layer, which is also evident from \cref{fig:vqe_circuit}. The MBVQE thus has 25 parameters and require 29 qubits. This shows that MBQC has the possibility of decreasing necessary depth by using the already present high entangling in MBQC-natural gates. Importantly this implies that the graph state can be kept fairly small and still produce interesting results.

\subsection*{Two-dimensional Heisenberg model}
\begin{figure}
    \centering
    \includegraphics[width=\linewidth]{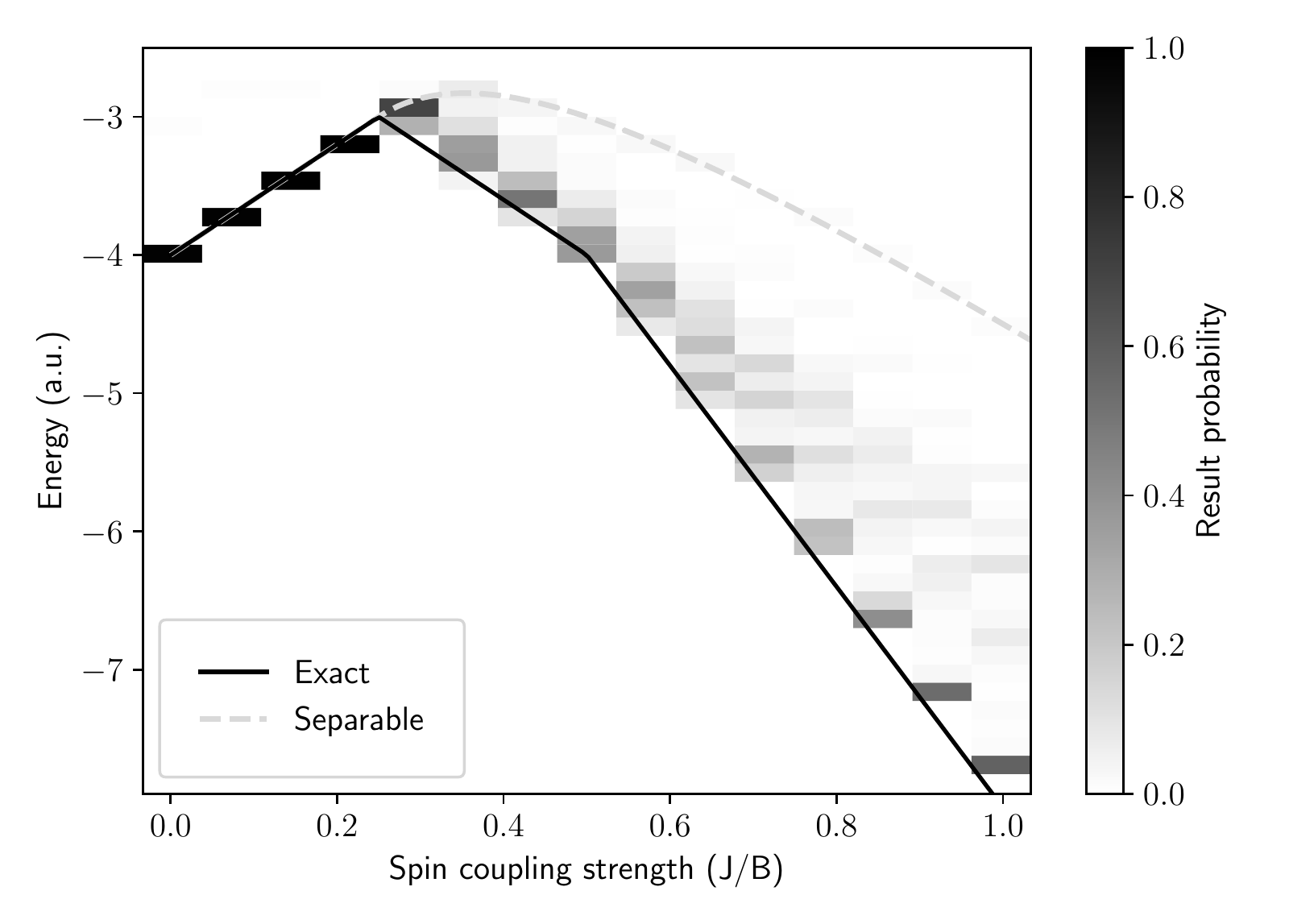}
    \caption{MBVQE results for the ground state of the $2 \times 2$ spin-lattice in an external homogeneous magnetic field woth nearest neighbour interactions. A total of 1020 simulations have been done for each of 15 different magnetic field strengths.}
    \label{fig:spin_lattice}
\end{figure}

Another interesting benchmark problem is the two-dimensional Heisenberg model. The problem is interesting because ground states are known to have a range of different degrees of entanglement for different regimes \cite{kandala_hardware-efficient_2017}. To test our methods and the MBVQE, we investigate the $2 \times 2$ lattice with nearest neighbour interactions. The qubit Hamiltonian representing the system is obtained by direct encoding, one qubit for each vertex in the lattice
\begin{equation}
    H_{spin} = B \cdot \sum_i Z_i + J \cdot \sum_{\langle i, j \rangle} \left(X_i X_j + Y_i Y_j + Z_i Z_j\right),
\end{equation}
where $\langle i, j \rangle$ runs over all nearest neighbours. The plot of \cref{fig:spin_lattice} is the results of 1020 independent MBVQE simulations for each of 15 different values of $J/B$. In the case of the Heisenberg model, we have opted for a depth-2 MBVQE, as a single layer has proven inadequate at producing enough entanglement in the weak field regime. With two layers of \cref{fig:vqe_circuit}, the MBVQE require 42 qubits (from \cref{eq:qubit_count} with $n=4$) and has a total of 38 variational parameters.

In the strong magnetic field approximation $J/B \leq 0.2$, it is clear that the algorithm finds the minimum with certainty. The reason is that the ground state in this regime is perfectly separable and thus is exactly obtainable from single-qubit rotations. The MBVQE accordingly finds that the parametrised entanglement is optimal for $R_{Z^{\tensor n}}(\theta=0/\pi)$. For decreasing magnetic field the ground state is increasingly entangled, and the optimal parameters are less trivial. This is also apparent from the vertical spread of the probability distributions in the MBVQE results for intermediate and weak magnetic fields $J/B \geq 0.2$. Notice, however, that the ground state energy is obtainable in all cases. We have observed that increasing the maximally allowed iteration count significantly reduces the vertical spread of the probability distributions and that the probability of obtaining the true ground state energy is almost exclusively a function of this parameter. This seems to suggest that the quantum part of the MBVQE is indeed highly effective at traversing the full Hilbert space.

\subsection*{Vehicle routing (Ising model)}
Finally, it has been demonstrated that solutions to NP-complete problems can be mapped onto ground states of Ising-type Hamiltonians \cite{lucas_ising_2014}. One extremely useful problem to which this applies is the Vehicle Routing problem \cite{feld_hybrid_2019}.

The problem Hamiltonian is given in terms of decision operators $x_{i, t}$ which is $=1$ if vertex $i$ is visited at time $t$, as well as the costs/weights $w_e$ associated with travelling along edge $e$. The Hamiltonian is then given by
\begin{equation}
    H = H_C + A \cdot H_P
\end{equation}
where the cost of a path is determined by
\begin{equation}
    H_C = \sum_{u,v\in V} w_{u, v} \sum_{t = 1}^n x_{u, t} x_{v, t+1}
\end{equation}
and the penalty term
\begin{equation}
    H_P = \sum_{v \in V}\Big( 1 - \sum_{t=1}^n x_{v, t} \Big)^2 + \sum_{t=1}^n\Big( 1 - \sum_{v \in V} x_{v, t} \Big)^2.
\end{equation}
makes sure that that path is indeed a Hamiltonian cycle. The scale factor $A$ applied to the penalty term must be chosen such that $A > \max_{e \in E} w_e$ for that to be the case. The number of variables is reduced by choosing some vertex to always be the first and last vertex in the cycle and fixing the corresponding $x_{i,t}$'s.

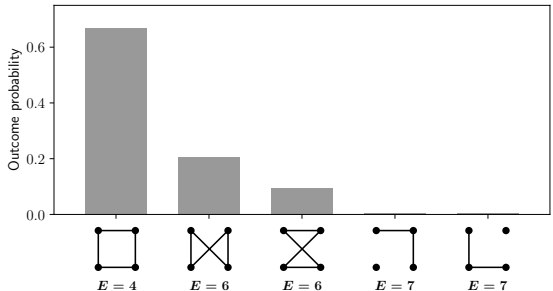
\begin{figure}
    \centering
    \resizebox{\linewidth}{!}{
    \begin{tikzpicture}
        \node at (0, 0) {\includegraphics{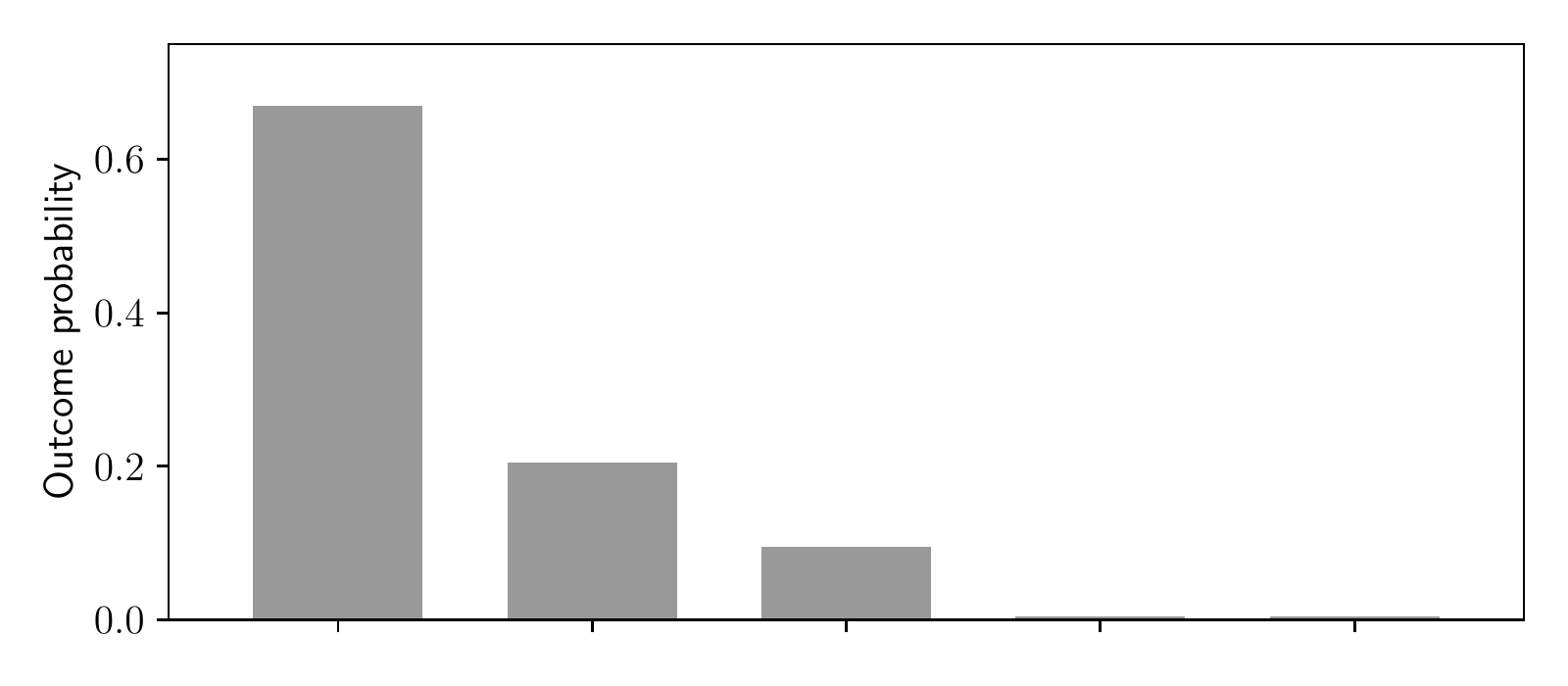}};
        \node[fill=none, circle] (center) at (0.65, -3.2) {};
        \node at ($(center) + (0, -1.2)$) {\scalebox{1.05}{\begin{tikzpicture}[x=1cm, y=-1cm, baseline={(current bounding box.center)}]
            \tikzstyle{every node} = [fill, circle, scale=0.6, font=\huge\boldmath];
            \draw[very thick] (0, 0) node (a) {} 
                           -- +(1, 0) node () {} 
                           -- +(1, 1) node () {} 
                           -- +(0, 1) node () {}
                           -- +(0, 0);
            \node[fill=none] at ($(a) + (0.5, 1.5)$) {$E=4$};
            \draw[very thick] ($(a) + (2.5, 0)$) node (a) {} 
                           -- +(1, 1) node () {} 
                           -- +(1, 0) node () {} 
                           -- +(0, 1) node () {}
                           -- +(0, 0);
            \node[fill=none] at ($(a) + (0.5, 1.5)$) {$E=6$};
            \draw[very thick] ($(a) + (2.5, 0)$) node (a) {} 
                           -- +(1, 1) node () {} 
                           -- +(0, 1) node () {} 
                           -- +(1, 0) node () {}
                           -- +(0, 0);
            \node[fill=none] at ($(a) + (0.5, 1.5)$) {$E=6$};
            \draw[very thick] ($(a) + (2.5, 0)$) node (a) {} 
                           -- +(1, 0) node () {} 
                           -- +(1, 1) node () {};
                           \node at ($(a) + (0, 1)$) {};
            \node[fill=none] at ($(a) + (0.5, 1.5)$) {$E=7$};
            \draw[very thick] ($(a) + (2.5, 0)$) node (a) {} 
                           -- +(0, 1) node () {} 
                           -- +(1, 1) node () {};
                           \node at ($(a) + (1, 0)$) {};
            \node[fill=none] at ($(a) + (0.5, 1.5)$) {$E=7$};
        \end{tikzpicture}}};
    \end{tikzpicture}
    }
    \caption{Simulated MBVQE results for the vehicle routing problem on the complete 4-vertex graph. For the penalty scale factor, $A=2.5$ was chosen. The algorithm was simulated a total of 100 times with random initial parameters. The bars in the figure show the total probability for measuring the corresponding bit strings averaged over all simulations. Only the five most probable outcomes are included. Each path is also presented together with the corresponding energy.}
    \label{fig:vehicle routing}
\end{figure}

The results of 100 independent MBVQEs applied to the vehicle routing Hamiltonian are presented in \cref{fig:vehicle routing}. The graph that we consider is the complete 4-vertex lattice (problem size $n=4$) with squared euclidean distance. That is, the cost of diagonal edges is 2 whereas the cost of the outer edges is 1. Also, we have chosen $A=2.5$. Each bar in the histogram is the average state probability determined over all runs. Only the five most probable outcomes are given. The MBVQE used to obtain the results is a single layer with $R_{Z^{\tensor n}}$ entangling, where the problem size is $n=9$, and thus has (\cref{eq:qubit_count}) $55$ parameters and require a corresponding $64$ qubits.

From the figure, it can be concluded that with a probability of about $2/3$ the true minimal solution is obtained. However, since we are seeking the minimal energy solution this doesn't have to be 100\%, as long as only a small number of samples are needed for that particular solution to appear. These results suggest that, on average, this will this will happen within two runs.

It is peculiar that the two Hamiltonian cycles with energy $E=6$ do not appear to be equally probable. This is unexpected since the paths appear entirely symmetric to the Hamiltonian. This reason is most probably due to the small sample size. However, another possible explanation is that parameters might be optimised in the order they are stored in computer memory which introduces a slight asymmetry in the explored Hilbert space. Also, the results were found to be highly dependent on the specific choice of the penalty factor $A$. But all in all, we can conclude that the MBVQE algorithm almost always results in a valid solution and that the minimal solution can be determined in only a small number of samples.
\section{Comparison to gate-based VQE}
We have so far discussed how to perform a measurement-based implementation of the VQE algorithm and seen how the entangling gate $R_{Z^{\tensor n}}$ performs for three particular problems. In this section we will discuss how the presented MBVQE compares to the standard gate-based VQE. 

First and foremost, it is important to note that the MBVQE algorithm presented in this paper (bottom half of \cref{fig:vqe_circuit}) is a direct implementation of its gate-based representation (top half of \cref{fig:vqe_circuit}), and the two produce, in the absence of errors and noise, entirely identical results. Since we are not considering any physical error models in the present work, all the presented results are independent of whether a gate-based VQE or the MBVQE was used. For a hardware-comparison of the two approaches, the most instructive comparison is that of time scaling or time complexity. 

If we assume an efficient technique for producing resource states, the bottleneck in optics is the measurement rate. There are two major factors that limit the measure rate: 1) measurement duration times including down time between measurements and 2) feed forward of measurement outcomes for adaptive measurements. It has been shown that measurement rates in the 1-10GHz are feasible \cite{inoue_toward_2023}. Currently, however, it is the process of feed forward that is the limiting factor. It is believed that it should be possible to keep up a feed forward process in the 10 MHz regime, although this is not a fundamental limit \cite{jonas_private}. Consequently, the execution time of an MBQC will scale linearly with the number of measurements. The number of measurements roughly equals the number of parameters in the VQE, and in our implementation (see \cref{eq:qubit_count}), this number scales as $O(3dn)$, where $d$ is the depth and $n$ the problem size. Therefore, the time-complexity of the optical MBVQE is $O(3dn)$ in the order of 10MHz.

In circuit QED the single most significant bottleneck is gate execution. Typical gate durations for single-qubit operations for trapped ions are of the order 100kHz and for two-qubit operations ($CNOT$) about 1kHz \cite{sonialopezbravo_ionq_2023}. Quite generally in circuit QED it is the two-qubit gates that are expensive. For transmons, two-qubit gates are executed in the order of 1-40MHz \cite{kandala_demonstration_2021, krantz_quantum_2019}. Note, that these figures are limited by the energy of the two-level systems and the relevant coupling strengths \cite{levitin_fundamental_2009, krantz_quantum_2019}. In our implementation of the VQE algorithm the only multi-qubit gate is the $R_{Z^{\tensor n}}$ which decomposes into $n$ $CNOT$ gates, a single $Z$ rotation, and finally another $n$ $CNOT$ gates. Thus, with one entangling step per layer, the time complexity is $O(2dn)$ ranging in the order of kHz to MHz.

Finally we note that the number of runs of a VQE algorithm required to obtain good estimates of operator expectation values is the same for both methods. Since each run of optical MBQC is about as fast to 10 times faster than that of circuit QED, the optical method will be significantly faster in the end. However, the two methods are still fundamentally different in their physical implementations. The errors that the systems exhibit will therefore be very different, leading to different advantages and obstacles which must be addressed.

\section{Conclusion and outlook}

We have presented the generic structure of our MBQC compiler that computes the measurement-based implementation of circuit model quantum algorithms. On top of direct translation, the compiler includes a module for reducing the MBQC by simulating all Clifford parts of the computation classically. We considered how a VQE ansatz can be prepared by the means of an MBQC, and in particular how to implement the algorithm onto a small resource state. Using tensor networks for simulating the MBVQE we have considered three types of Hamiltonians, namely molecular, Heisenberg models, and Ising type.

We have concluded, that problems of a practical matter can be solved within MBQC using only reasonably small resource states. These are promising prospects for developing measurement-based platforms for quantum computing that can solve problems outside the scope of boson sampling.

We also note that developing methods for efficiently traversing the graph LC-equivalence classes will be important in order to determine optimal resource states for use with MBQC. Such methods will allow for engineering MBQCs tailored to the resource states one has available, and will thus significantly increase the range of realisable algorithms. This will be subject to future work.

\section*{Acknowledgements}
\begin{itemize}[leftmargin=0pt ,label={}]
    \item We would like to thank the PhotoQ consortium funded by the Innovation Fund Denmark for numerous discussions, including Mark N. Jones and Kaur Kristjuhan at MQS, and Jonas Neergaard-Nielsen, Jens A. H. Nielsen, Emil Østergaard, and Abhinav Verma at DTU.
    \item Stig E. Rasmussen and Mogens Dalgaard of Kvantify ApS. for being great sources of information on things concerning variational quantum algorithms as well as invaluable feedback on the manuscript.
    \item We also acknowledge discussions with Peter Lodahl, Stefano Paesani, and Anders Sørensen at the Center for Hybrid Quantum Networks (Hy-Q) at the Niels Bohr Institute during the early stages of this work.
    \item The bulk of the simulations done for this article was performed on the UCloud interactive HPC system, which is managed by the eScience Center at the University of Southern Denmark.
\end{itemize}

\printbibliography

\end{document}